\documentclass[twocolumn,showpacs,preprintnumbers,amsmath,amssymb,prb]{revtex4}
\newcommand{\be}{\begin{equation}}
\newcommand{\ee}{\end{equation}}
\newcommand{\bea}{\begin{eqnarray}}
\newcommand{\eea}{\end{eqnarray}}
\usepackage{graphicx}
\usepackage{float,afterpage}
\usepackage{dcolumn}
\usepackage{bm}
\begin{document}
\title{A quantum mechanical derivation of the Schwarzschild radius and
its quantum correction using a model density distribution: Skin of a
black hole}
\author{Subodha Mishra}
\affiliation {Department of Physics \& Astronomy,
 University of Missouri, Columbia, MO 65211, USA}
\date{\today}
\begin{abstract}
Using a single particle density distribution for a system of 
self-gravitating particles which ultimately forms a black hole, we from a 
condensed matter point of view derive the Schwarzschild radius and 
by including the quantum mechanical exchange energy we find a small 
correction 
to the Schwarzschild radius, which we designate as the skin of the black 
hole.
\pacs{PACS:  }
\end{abstract}
\maketitle


\section{Introduction}
In the evolution of a star, when the radius of a star becomes less than a 
certain limit, called the Schwarzschild radius $R_{Sch}=2GM/c^2$, 
where M is the mass of the object; $G$, Newton's universal 
gravitational constant, and $c$, velocity of light; neither 
light nor material particles can escape from the star. Thus, one finds 
from above, the Schwarzschild radius corresponds to the situation when the 
escape velocity is equal to the velocity of light, a particle (or a 
photon) coming from a large distance when passes near by  the black 
hole\cite{raine,wheel}, 
it is not only attracted towards it but also its orbit diverges from a 
straight line.It may so happen that if  the particle goes too close to a 
black hole it is likely to be trapped and hence, it cannot escape to 
infinity. As a special case, if the particle travels straight to the 
center of the black hole, it falls inside it and is lost forever.
An interesting 
problem that is associated with the formation of a black hole is the final 
collapse of a massive star. This happens when the nuclear fuel inside the  
central core of the star gets exhausted. At this stage, it is the dominance 
of the gravitational attraction among the particles within a star over the 
internal outward pressure which makes the star to collapse. A black hole,
 though may be formed from baryons, leptons etc; the exterior observer 
cannot 
have access to the details of the inside of the black hole. The observer 
probes the black hole mass M, electromagnetic charge Q and angular 
momentum J. This is refered to as baldness of the black hole or as Wheeler 
describes, a black hole has no hair\cite{wheel}. But actually it has only 
three hairs, 
M,\ Q,\ J. We in this paper derive the Schwarzschild radius of a system of 
gravitating particles quantum mechanically and show that when quantum 
exchange correction is taken into account, there is a thin 
correction to this radius, which we designate as the skin of the black 
hole.
It has been shown previously\cite{duff} that using the quantum field 
theoretic method, by
including a single-closed-loop in the self energy, a quantum 
correction to
the classical Schwarzschild solution of the order of $\sim G^2$ can be 
found. This
comes from the gravity sector. The correction that we find, comes from the 
exchange part of the matter-energy sector of the black hole. Our correction is 
also of the 
order of $\sim G^2$.

We here consider a black hole with mass only, a Schwarzchild\cite{raine} 
type black 
hole. As 
described  above, since the black hole forms out of particles due to 
immense gravitational attraction, we consider the system of particles as a 
many particle system that ultimately forms a black hole. 
We have succeeded in developing a quantum mechanical
approach\cite{sm1,sm2} for a system of self-gravitating particles using
Newtonian gravity. We calculate the 
energy content of the system including exchange interaction. In order to 
calculate the total energy of a star, we choose 
a trial single-particle density to account for the distribution of 
particles within star. The form of our single-particle density is such 
that it has a singularity at the origin. Applying it to the case of a 
neutron star, we not only arrive at a compact expression for the radius of 
the neutron star, but also obtain an expression for the binding energy of 
the star which varies with the particle number\cite{levy} as $N^{7/3}$, 
where N is 
the particle number. Such a dependence with $N$ is in agreement with those 
of the earlier workers.\cite{levy} The aim of the present work is to give 
a 
derivation of the socalled Schwarzschild radius even without using GTR and 
relativistic quantum mechanics. By accounting for the exchange effects due 
to the interparticle correlations to the total ground state energy of the 
system, we find a quantum correction to the Schwarzschild radius.

 \section{Mathematical formulation}

In order to describe a system of N self-gravitating particles in absence 
of any source for radiation, we use a Hamiltonian of the form:
\be
H=\sum_{i=1}^N\frac{-\hbar^2\nabla_i^2}{2m}+\frac{1}{2}\sum_{i=1}^N
\sum_{j=1,i\ne j}^Nv(|\vec X_i-\vec X_j|),
\ee
where $v(|\vec X_i-\vec X_j|)=-g^2/|\vec X_i-\vec X_j|$, is the 
interparticle interaction between  pair of gravitating particles 
and $g^2=Gm^2$, 
$m$ being the mass of particle and G being the universal gravitational 
constant. In the present case we confine ourselves to the system of 
neutrons only. Since the wavefunction of a neutron star is not known, we 
proceed\cite{sm1,sm2} to evaluate the total kinetic energy of the system 
using a model 
density distribution function for particles (neutrons) within the neutron 
star. The particles being fermions, we use the Thomas-Fermi formula for 
calculating the total kinetic energy of the system. For an infinite 
many-fermion system, the average particle density and the fermi momentum 
of 
a particle are related to each other as:
\be
n=\frac{k_F^3}{3\pi^2}.
\ee
For a finite system like  the star, since the density distribution is a 
function of radius vector $\vec r$, the fermi energy of a particle is 
supposed to be dependent on $r$. For an infinite many-fermion system, the  
total kinetic energy of the system\cite{fetter} is given as 
\bea
<KE>_{inf}=\frac{3}{5}n\epsilon_F=\frac{3}{5}n(3\pi^2n)^{2/3}\frac{\hbar^2}{2m}
\nonumber\\
=\frac{3\hbar^2}{10m}(3\pi^2)^{2/3}n^{5/3}\label{ke},
\eea
In analogy with 
Eq.(\ref{ke}), 
for a finite system\cite{land}, we write,
\be
<KE>=\frac{3\hbar^2}{10m}(3\pi^2)^{2/3}\int d\vec r [\rho(\vec 
r)]^{5/3}\label{kee},
\ee
The total energy E of the system is given as 
\be
E=<H>=<KE>+<PE>,
\ee
where
\be
<PE>=-\frac{g^2}{2}\int d\vec r d\vec r'\frac{\rho(\vec r)\rho(\vec r')}
{|\vec r-\vec r'|}\label{pe}
\ee
The expression given in Eq.(\ref{pe}) is written in the Hartree 
approximation. In order to find E, we choose a trial single-particle 
density of the form
\be
\rho(\vec r)=A\frac{exp[-(\frac{r}{\lambda})^{1/2}]}
{(\frac{r}{\lambda})^{3/2}}\label{den}
\ee
where A is the normalization constant, which is determined using the 
relation
\be
\int\rho(\vec r)d\vec r=N .
\ee
As one can see from Eq.(\ref{den}), $\rho(\vec r)$ is singular at $r=0$. 
In general one could choose a single particle density of the form
\be
\rho(\vec r)=A\frac{exp[-(\frac{r}{\lambda})^{\nu}]}
{(\frac{r}{\lambda})^{3\nu}},\label{deng}
\ee
where $\nu=1,2,3,4...or\ \frac{1}{2},\frac{1}{3},\frac{1}{4},...$. Integer 
values of $\nu$ are not permissible because they make the normalization 
constant infinite. Out of the fractional values, $\nu=\frac{1}{2}$ is 
found to be most appropriate, because, as we shall see later, it gives 
the expected upper limit for the critical mass of a neutron 
star\cite{sm1}, beyond which black hole formation takes place. Any other 
value of $\nu$ would give rise to a different value for the critical mass. 
Also because if $\nu$ goes to zero (like $1/n$, $n\rightarrow\infty$), 
$\rho(r)$ would tend to the case of a constant density as found in an 
infinite many-fermion system. Having accepted the value $\nu=\frac{1}{2}$, 
the parameter $\lambda$ associated with $\rho(r)$ is determined after 
minimizing $E(\lambda)=<H>$ with respect to $\lambda$. This is how, we are 
able to find the total energy of the system corresponding to its lowest 
energy state.
After evaluating the integral shown in Eq.(\ref{kee}) and Eq.(\ref{pe}), 
we 
obtain
\be
E(\lambda)=\frac{\hbar^2}{m}\frac{12}{25\pi}(\frac{3\pi 
N}{16})^{5/3}\frac{1}{\lambda^2}-\frac{g^2N^2}{16}\frac{1}{\lambda}\label{bige}
\ee
Differentiating this with respect to $\lambda$ and then equating it with 
zero, we obtain the value of $\lambda$  at which the minimum occurs. This 
is found as:
\be
\lambda_0=\frac{72}{25}\frac{\hbar^2}{mg^2}(\frac{3\pi}{16})^{2/3}
\frac{1}{N^{1/3}}\label{lam}
\ee
Here we are only concerned with the total kinetic energy of the system. At 
$\lambda=\lambda_0$, we have,
\be
<KE>=0.015441\frac{mg^4}{\hbar^2}N^{7/3}\label{kes}
\ee
The total energy E of the system is found to be just negative of this.
\section{Derivation of Schwarzschild Radius}

We 
now try to calculate the average velocity of a particle within the neutron 
star using Eq.(\ref{kes}). Let us denote it by $<\vec v^2>$. If M denotes 
the total mass of the neutron star, one writes
\be
<KE>=\frac{1}{2}M<\vec v^2>\label{ket},
\ee
where $M=Nm$. By comparing Eq.(\ref{ket}) with Eq.(\ref{kes}), one 
obtains,
\be
<\vec v^2>=0.030882\frac{g^4}{\hbar^2}N^{4/3}.
\ee
From the expression for the total kinetic energy of an infinite 
many-fermion system\cite{fetter}, one finds that the average velocity of a 
particle within the system is $\sim 0.77v_f$, $v_f$ being the fermi 
velocity\cite{fetter} of the particle within the system, which is maximum 
velocity of 
that particle. From this, one clearly sees that the maximum velocity of a 
particle belonging to an infinite many-fermion system is greater than the 
average particle velocity. In view of this fact, we could write the 
maximum velocity of a particle within a neutron star as
\be
<\vec v^2>_{max}=\alpha <v^2>=0.030882\alpha\frac{g^4}{\hbar^2}N^{4/3}
\ee
where $\alpha$ is a constant whose value is to be greater than unity and 
it is to be calculated later. $v_{max}$ can be identified as the escape 
velocity of a particle within a neutron star.
According to special theory of relativity, $\vec v^2_{max}$ is to be less 
than $c^2$, c being the velocity of light. That is,
\be
0.030882\alpha\frac{g^4}{\hbar^2}N^{4/3}\le c^2,
\ee
From this it follows that
\be
N\le\frac{13.574409}{\alpha^{3/4}}(\frac{\hbar 
c}{g^2})^{3/2}=N_c\ (say),\label{nu}
\ee
having $g^2=Gm^2$. Substituting Eq.(\ref{nu}) in Eq.(\ref{lam}), one 
finds 
that,
\be
\lambda_0\ge\lambda_c=\frac{Gm}{c^2}\alpha^{1/4}[0.8483718(\frac{\hbar 
c}{g^2})^{3/2}].
\ee
If we define the radius of a neutron star as $R_0=2\lambda_0$, we have 
the expression for the  critical radius as,
\be
R_c=2\lambda_c=2\frac{Gm}{c^2}\alpha^{1/4}[0.8483718(\frac{\hbar 
c}{g^2})^{3/2}]=\frac{2GM}{c^2}\label{rad}
\ee

Our identification about the radius $R$ of the star with $2\lambda_0$ is 
based on the use of socalled quantum mechanical tunneling\cite{karp} 
effect. Classically, it is well known that a particle has a turning 
point 
where the potential energy becomes equal to the total energy. Since the 
kinetic energy and therefore the velocity are equal to zero at such a 
point, the classical particle is expected to be turned around or reflected 
by the potential barrier. From the present theory it is seen that the 
turning point occurs at a distance $R=2\lambda_0$. This is the reason why 
we identify $2\lambda_0$ with the radius of a star. For $R>2\lambda_0$, 
a particle, belonging to the system, may have an access to the region beyond 
$R>2\lambda_0$, because of quantum mechanical tunneling, but is 
forbidden by classical theory. 

$R_c$ as given in Eq.(\ref{rad}) is being identified as 
the so called Schwarzschild radius which we have derived here by treating 
the system as a quantum many-body system. When $R_0\le R_c$, the 
corresponding 
neutron star becomes a black hole. From Eq.(\ref{nu}), we therefore find 
that the lowest mass of the neutron star beyond which black hole formation 
takes place is given as
\be
M_c=mN_c=\frac{13.574409}{\alpha^{3/4}}m(\frac{\hbar c}{g^2})^{3/2}
\ee
In order to determine $\alpha$, we now try to evaluate the limiting mass 
of a neutron star following the general expression for the radius of a 
star. Beyond this mass, the black hole formation is likely to take place. 
For that, we consider the situation when
\be
(R_0=2\lambda_0)= (R_{sch}=\frac{2GM}{c^2}),
\ee
where $M=Nm$. From this, we arrive at
\be
N\ge(1.696758)(\frac{\hbar c}{g^2})^{3/2}=N_c.\label{nc}
\ee
Since the expression in the right hand side of Eq.(\ref{nc}) should be 
equal to the one given in right hand side of Eq.(\ref{nu}), we must have 
$\alpha=16$. Under this situation, we have
\be
v^2_{max}=0.494112\frac{g^4}{\hbar^2}N_c^{4/3},
\ee
where $N_c=1.696758(\frac{\hbar c}{g^2})^{3/2}$, which, when evaluated, 
becomes $3.7390777\times 10^{57}$. For a neutron star in which the number 
of neutrons exceeds $N_c$, it has the tendency of forming a black hole. In 
that case, its mass must exceed $M=M_c=mN_c=3.12213\ M_{\odot}$, 
$M_{\odot}$ 
being the solar mass.

\section{Quantum Correction}

So far we have been discussing about the quantum mechanical derivation of 
the 
Schwarzschild radius $R_{Sch}$. The very form of $R_{Sch}$ shows that it 
is a classical result, leaving aside the fact the number of particles N 
within a neutron star\cite{wes} is to be less than $N_c$ where 
$N_c=1.70(\frac{\hbar 
c}{Gm_n^2})^{3/2}$, which involves the Planck's constant $\hbar$.
Now, inorder to account for the 
quantum corrections to $R_{Sch}$, we go beyond the Hartree 
contribution to the  total energy of the system. That is the exchange 
correction or Hartree-Fock(HF) term\cite{beth} over the Hartree result 
(direct contribution). Since 
the 
HF-correction term is non-local we make use of the local density 
approximation\cite{beth} to write it as,
\be
<PE>_{ex}=\frac{3}{2\pi}(3\pi^2)^{1/3}g^2\int d\vec r[\rho(\vec r)]^{4/3}.
\ee
This when evaluated gives
\be
<PE>_{ex}=\frac{27}{4}(\frac{1}{16\pi})^{4/3}(3\pi^2)^{1/3}g^2
\frac{N^{4/3}}{\lambda}.
\ee
With the inclusion of this extra term, the  expression for $E(\lambda)$, 
Eq.(\ref{bige}) is minimized with respect to $\lambda$ and we  arrive at
\be
\lambda_0'=\frac{72}{25}(\frac{3\pi}{16})^{2/3}\frac{\hbar^2}{mg^2}
\frac{1}{N^{1/3}}[1+\frac{1.8010}{N^{2/3}}]\label{lamor}
\ee
Following the  argument discussed earlier, we identify the radius of the 
neutron star by $R'_0=2\lambda'_0$. As before writing 
${v'}^2_{max}=16<{v'}^2>$ 
and using the condition that ${v'}^2_{max}\le c^2$, we  obtain
\be
N\le N'_c=1.696758(\frac{\hbar c}{g^2})^{3/2}[1+0.4747761(\frac{g^2}{\hbar 
c})]
\ee
Corresponding to $N'_c$ , the new  expression for the critical radius 
$R'_c$ becomes
\be
R'_c=2\lambda'_c=2\frac{GM_c}{c^2}[1+0.7912723(\frac{g^2}{\hbar 
c})]\label{rp} 
\ee
where $M_c=mN_c=1.696758(\frac{\hbar c}{g^2})^{3/2}$. The above 
expression, Eq. (\ref{rp}) is obtained by keeping  terms upto order 
$(\frac{g^2}{\hbar c})$ only in Eq. (\ref{lamor}). 
For $N>N'_c$, the neutron star is likely to go over the black hole stage. 
From Eq.(\ref{rp}), we find that the second term within the square 
bracket, forms 
the quantum correction to the Schwarzschild radius. As expected, it 
involves the gravitational fine structure constant $(\frac{g^2}{\hbar 
c})$. Since it is of the order $10^{-39}$,  obviously it makes  an 
extremely small correction to $R_{Sch}$. 
It has been shown earlier\cite{duff} that using the quantum field 
theoretic method and by 
including a single-closed-loop in the self energy, a quantum correction to 
the classical Schwarzschild solution of the order of $\sim G^2$ can be 
found. This 
comes from the gravity sector. The correction that we get is also of the 
order $\sim G^2$ but it comes from the 
exchange part of the matter-energy sector of the black hole.

\section{Conclusion}
We in this paper derive the Schwarzschild radius of a black hole from a 
condensed matter point of view by using a single particle density 
distribution for the many-body self-gravitating system which ultimately 
forms a black hole. By incorporating the quantum mechanical exchange 
interaction, we also find a thin correction to the Schwarzschild radius 
which we designate as the skin of the black hole.

We thank F. Hehl for critically reading the manuscript and bringing to our 
notice the Ref.3.


\begin{thebibliography}{99}
\bibitem{raine}D. Raine and E. Thomas, {\it Black Holes}, (Imperial 
College Press, London, 2005).
\bibitem{wheel}J. A. Wheeler and K. Ford, {\it Geons, black holes and 
quantum foam: A life in physics}, (W. W. Norton \& Company, 2000).
\bibitem{duff}M. J. Duff, Phys. Rev. D. {\bf 9},  1837 (1974).
\bibitem{sm1}D. N. Tripathy and S. Mishra, Int. J. Mod. Phys. {\bf D7}, 
431 (1998).
\bibitem{sm2}D. N. Tripathy and S. Mishra, Int. J. Mod. Phys. {\bf D7},
917 (1998).
\bibitem{levy}J. M. Levy-Leblond, J. Math. Phys. {\bf 10}, 806 (1969).
\bibitem{fetter}A. L. Fetter and J. D. Walecka, {\it Quantum theory of 
many-particle system}, (McGraw Hill, New York, 1971).
\bibitem{land}L. D. Landau and E. M. Lifshitz, {\it Quantum 
Mechanics}, (Pergamon Press, Oxford, 1965).
\bibitem{karp}M. Karplus and R. N. Porter, {\it Atoms and Molecules}, (W.
A. Benjamin Inc, California, 1970)
\bibitem{wes}P. S. Wesson, {\it Cosmology and Geophysics}, (Adam Hilger 
ltd, Bristol, 1978).
\bibitem{beth}H. A. Bethe and R. W. Jackiw, {\it Intermediate Quantum 
Mechanics},(W.A. Benjamin Inc, London, 1968).
\bibitem{schiff}L. I. Schiff, {\it Quantum Mechanics}, (McGraw Hill, 
Singapore, 1985).
\end{thebibliography}
\end{document}